\documentclass[12pt]{article}
\usepackage{graphicx}
\usepackage{amsmath}
\usepackage{amssymb}
\usepackage{natbib}
\usepackage[matrix,arrow]{xy}
\usepackage{lscape}
\usepackage{subfigure}
\usepackage{epstopdf}
\usepackage{bm}
\usepackage{moreverb}
\usepackage{setspace}
\usepackage{lineno}
\usepackage{xcolor}

\title{A data-driven method for the stochastic parametrisation of subgrid-scale tropical convective area fraction}
\author{G. A. Gottwald\thanks{School of Mathematics and Statistics, University of Sydney, NSW 2006, Australia; georg.gottwald@sydney.edu.au} \and K. Peters\thanks{ARC Centre of Excellence for Climate System Science, School of Mathematical Sciences, Monash University, Clayton, VIC 3800, Australia; karsten.peters@monash.edu.au; now at Max Planck Institute for Meteorology, Hamburg, Germany} \and  L. Davies\thanks{School of Earth Sciences, University of Melbourne,VIC 3010, Australia}}

\begin{document}

\maketitle

\begin{abstract}  
Observations of tropical convection from precipitation radar and the concurring large-scale atmospheric state at two locations (Darwin and Kwajalein) are used to establish effective stochastic models to parameterise subgrid-scale tropical convective activity. Two approaches are presented which rely on the assumption that tropical convection induces a stationary equilibrium distribution. In the first approach we parameterise convection variables such as convective area fraction as an instantaneous random realisation conditioned on the large-scale vertical velocities according to a probability density function estimated from the observations. In the second approach convection variables are generated in a Markov process conditioned on the large-scale vertical velocity, allowing for non-trivial temporal correlations. Despite the different prevalent atmospheric and oceanic regimes at the two locations, with Kwajalein being exposed to a purely oceanic weather regime and Darwin exhibiting land-sea interaction, we establish that the empirical measure for the convective variables conditioned on large-scale mid-level vertical velocities for the two locations are close. This allows us to train the stochastic models at one location and then generate time series of convective activity at the other location. The proposed stochastic subgrid-scale models adequately reproduce the statistics of the observed convective variables and we discuss how they may be used in future scale-independent mass-flux convection parameterisations.
\end{abstract}

{\bf{keywords: }}{tropical convection, stochastic parameterisation; convective parameterisation; general circulation model; precipitation radar; cloud base mass flux}

\maketitle



\section{Introduction}
\label{sec:Intro}
Despite a remarkable increase in complexity and resolution of general circulation models (GCMs), the representation of deep convection, which ultimately serves to drive the general circulation, is still associated with large uncertainties \citep{Flato13}. 
The inadequate representation of atmospheric convection in GCMs is responsible for considerable uncertainty in estimating climate sensitivity \cite[][and references therein]{BonyEtAl15} and ambiguities in the numerical simulation of the Earth's climate, for example when comparing the inter-model mean and spread of hydrological-cycle related variables of the CMIP5 ensemble to observations \cite[e.g.][]{Jiang12,Tian13,Lauer13}.
An improved representation of fundamental atmospheric processes, such as convection, is therefore considered to be of utmost priority in the model design \citep{Stevens13,Jakob14}. 

Atmospheric convection cannot be resolved by the model grid of GCMs currently used for climate projections and must therefore be parameterised. More than four decades ago, the pioneering works of \cite{Ooyama64} and \cite{Manabe65} laid the foundations for the development of increasingly complex convective parameterisation schemes (see \cite{Arakawa04} for a review and \cite{Randall13} for an outlook). As a result of this development, GCMs are now capable of reliably capturing the overall amount of precipitation. However, spatial distributions and variance often compare poorly to observations \cite[e.g.][]{Dai06,Pincus08,Stephens10}. Further, capturing the statistical relationship between convective activity and the large-scale environment
is a challenging task not often met by current GCMs. For example, \cite{Holloway12} show that a model with parameterised convection does not adequately reproduce the relationship between convective activity and vertical pressure velocity $\omega$ as found in a cloud-system resolving model with explicit convection. Using the observational datasets used in this study (cf. Section \ref{sec:Data}), preliminary analysis of the relationship between rain rates and $\omega$ at 500 hPa ($\omega_{500}$) in a state-of-the art climate model \cite[ECHAM6.2, c.f.][for a model description]{Stevens13b} over Darwin and Kwajalein yield similar negative results, with the relationship being qualitatively better captured over Kwajalein (not shown). 

Conventional convective parameterisations tend to be of a deterministic nature and represent only the mean effect of the small-scale unresolved convective processes on the resolved large-scale environment on the scale of the numerical grid. In these parameterisations, it is assumed that for any given resolved large-scale state of the atmosphere-ocean system there exists a single possible response at the small-scale convective state feeding back upon the large-scale state. There is, however, a mounting body of evidence that actual observed convection does not obey deterministic relationships between large-scale variables and convective scales \cite[e.g.][]{Peppler89,Sherwood99,Holloway09,Stechmann11,Davies13,Peters13}. Furthermore cloud-resolving models (CRMs) reveal a high degree of variability of small-scale convective activity for a given large-scale state. This challenges the usefulness of employing deterministic relationships between convective activity and large-scale variables \citep{XuEtAl92,CohenCraig06,Shutts07}. 

The complex chaotic dynamics of small-scale processes is widely recognised to give rise to the observed variability. For example, \cite{Hohenegger06}, using an ensemble of limited-area convection permitting simulations over the European Alps, identified gravity waves generated in regions of diabatic forcing (i.e.~moist convection) as the main source of error growth in their simulations. A lack of variability in the high-frequency, small-scale convective processes can dynamically propagate upscale and cause GCMs to misrepresent low-frequency large-scale variability \citep{RicciardulliGarcia00,HorinouchiEtAl03}. Model simulations and observations suggest that a stochastic approach to subgrid-scale parameterisations is needed \citep{Palmer01,Palmer12}. The recent increase of resolution of the numerical cores adds to the failure of purely deterministic parameterisations: For example, numerical square grids with edge lengths $\mathcal{O}(100\mbox{km})$ and less do not contain sufficient cumulus clouds to allow for the estimation of meaningful averages \citep{PalmerWilliams08}, and there is a need for a stochastic resolution aware parameterisation \citep{Arakawa11,Arakawa13}.\\

A plethora of stochastic subgrid-scale parameterisations for convection have been developed. \cite{Buizza99} applied random perturbations to the parameterised tendencies in the operational ECMWF Integrated Forecast System (IFS) improving its forecast skill. \cite{Lin00,Lin03} introduced random perturbations to convective available potential energy (CAPE) and to the heating profile of the host convective scheme improving on the statistics of tropical intraseasonal variability. \cite{Bright02} introduced random perturbations to the trigger function of the \cite{KainFritsch90} convection scheme, and \cite{Teixeira08} randomly perturbed tendencies from a deterministic convection scheme by sampling from a normal distribution. \cite{Plant08} used  random samples of a distribution of convective plumes to match a required grid-box mean convective mass flux.  
Their scheme has been successfully applied to a limited area model-ensemble over central Europe \citep{Groenemeijer12}. \cite{BernerEtAl05} used ideas from cellular automata to introduce stochastic forcing to the streamfunction to model the effect of mesocale convective systems. \cite{Bengtsson13} developed a stochastic convective parameterisation based on cellular automata via a moisture convergence closure, and showed that in a limited area model-ensemble framework over Scandinavia, the parameterisation leads to a desired increase in spread of the resolved wind field in regions of enhanced deep convection. \cite{Majda02} and \cite{Khouider03} drove a mass-flux convective parameterisation with a stochastic model based on convective inhibition. \cite{Khouider10} developed the stochastic multi-cloud model (SMCM) evolving a cloud population consisting of three cloud types associated with tropical convection (congestus, deep convective and stratiform clouds) by means of a Markovian process conditioned on the atmospheric large-scale state. This model has been shown to adequately simulate tropical convection and associated wave features in a simple two-layer atmospheric model \cite[e.g.][]{Frenkel12,Frenkel13,DengEtAl15} and to reproduce observed convective behaviour when observation-based transition time scales between cloud-types are adopted \citep{Peters13}. For a more comprehensive review on current stochastic subgrid-scale parameterisations of convection see, for example, \cite{NeelinEtAl08} and \cite{PalmerWilliamsBook10}.

Despite successfully capturing the observed high-frequency variability stochastic subgrid-scale parameterisations are often difficult to tune and very sensitive to the choice of the parameters as shown for example by \citet{Lin00,LinNeelin02,Lin03}. There has, however, not been much effort in alleviating this difficulty by imposing observational constraints on the parameterisation. The limited availability of high-quality, long-term datasets of concurring large-scale and convective scale observations surely contributes to this omission. We list recent works in that direction. \cite{NeelinEtAl08} and \cite{Stechmann11} used observed relationships between column integrated water vapour and precipitation to inform a physics-based stochastic model to simulate the onset and duration of very strong convection. \cite{Horenko11} developed a framework which allows for a purely data-based Markov chain parameterisation allowing for nonstationary data to model cloud cover. \citet{DeLaChevrotiereEtAl14} used data to infer the transition rates used in the SMCM by employing a Bayesian framework. \cite{Dorrestijn13} used data from large-eddy simulations to design a data-driven multi-cloud model. The transitions between different cloud types are calculated using Markov chains which are conditioned on large-scale variables. 
More recently \cite{Dorrestijn15} have successfully employed that model on observational data obtained in Darwin.

We complement here the suite of data-driven stochastic models of tropical convection by using observations to build a simple entirely observation-based stochastic model. An entirely observation-based model lacks the transparency of physics-based models, but is potentially more accurate. We exploit available long-term observations of the large-scale atmospheric and the concurring small-scale convective state over Darwin and Kwajalein \citep{Davies13}. The observations are used to inform stochastic models for the convective area fraction (${\rm CAF}$) and the rain rate. We present two stochastic models. In the first model, ${\rm CAF}$ (or the rain rate) is treated as an uncorrelated random variable conditioned on the large-scale vertical motion $\omega_{500}$. To incorporate non-trivial temporal correlations, we propose a second stochastic model whereby ${\rm CAF}$ (or the rain rate) are modelled as a Markov chain conditioned on $\omega_{500}$. The stochastic parameterisations can be constructed at either location and then be applied to observations of large-scale variables from the respective other location. Despite the different atmospheric and oceanic conditions of the two geographical locations, the stochastic models reproduce the observed statistics of the convective activity such as mean, variance and skewness. 

The underlying premise of our approach is that the stationary stochastic process relating small-scale convective activity and large-scale vertical velocity is sufficiently universal in the sense that the stochastic model can be transferred from one geographical location to another one. Using a Kullback-Leibler information criterion for the conditional probabilities of convective activity as well as quantile regression for the observational data we establish that for the two regions considered here, it is sufficient to correct for the large-scale variables by a simple linear translation to account for the respective ambient atmospheric and oceanic regimes at Darwin and Kwajalein.
It turns out that for mid-level vertical velocities no translation is required and one can apply the model trained at Kwajalein (Darwin) directly to data in Darwin (Kwajalein).

Although most stochastic parameterisations involve CAPE, we follow \cite{Davies13}, \cite{Peters13} and \cite{Dorrestijn15} and relate the observed convective state to $\omega_{500}$. \cite{Dorrestijn15} find that convection is highly correlated with column-integrated vertical velocity starting several hours before the onset of deep convection. This is not surprising as large-scale vertical motion in the tropics is directly related to deep convection. Conditioning convective states on vertical motion raises the question of cause-and-effect ambiguities \cite[see e.g.~][for a discussion]{Arakawa04,Peters13}. On the one hand, convection induces large-scale ascending motion through latent heating, which then facilitates further convection. On the other hand, pre-existing large-scale ascending motion (or convergence) may facilitate the development of convection \citep{Hohenegger13,Birch14} which then further increases large-scale ascending motion. We thus argue that tropical convection and large-scale ascending motion are intimately linked via a positive feedback loop, limited by the available energy in the atmospheric column and its close environment. We stress that the stochastic parameterisation we propose does not rely on nor presume any cause-and-effect relationship between vertical velocities and convective activity such as ${\rm CAF}$. The models only utilise observed statistical relationships such as conditional probabilities and transition probabilities.\par

We use ${\rm CAF}$ (as well as rain rate data) to characterise convective activity (cf. \cite{Dorrestijn13} and \cite{Bengtsson13}). Our motivation to formulate the parameterisation with respect to ${\rm CAF}$ is that it can be used to close convection schemes since measures of convective activity such as precipitation are linearly related to the area covered by the precipitation feature \citep{Craig96,Nuijens09,YanoPlant12b,Davies13}.\\ Furthermore, parameterisations for ${\rm CAF}$ can be by construction included in the framework of resolution independent parameterisations \citep{Arakawa11,Arakawa13,Wu14}. Current mass-flux convection schemes used in operational GCMs assume the area covered by convective updrafts to be negligible compared to the cloud-free part of a model grid box -- the so-called assumption of ``scale-separation''. 
This assumption breaks down once the resolution of the GCM becomes high enough such that the area covered by convective updrafts can occupy large parts of or even an entire grid box. Parameterisations for ${\rm CAF}$ are naturally scalable and could be used to mitigate this problem \citep{Arakawa13,Wu14}. Furthermore, most currently employed schemes are mass-flux schemes and need to predict the vertical mass flux at cloud base. The mass flux at cloud base could be determined by explicitly assigning an area to the convective updraft together with an updraft velocity. The effect of convection on the environment could be implemented by formulating the dependency of the vertical eddy fluxes of thermodynamic variables on updraft fraction as defined by \cite{Arakawa13} and \cite{Wu14} or through allowing convectively induced subsidence impact on neighbouring grid boxes \citep{Grell14}. Although using ${\rm CAF}$ allows for a certain scale-adaptivity, an increase in resolution would prohibit to identify the grid-box state as the large-scale environment. In this case, defining the large-scale environment as the average over a number of surrounding grid-boxes could be used \cite[e.g.][]{Keane12}.\\

The paper is organised as follows. We introduce the observational datasets along with a comparison of convective behaviour in Darwin and Kwajalein in Section \ref{sec:Data}. We then use the data to construct the stochastic subgrid-scale convection parameterisations in Section \ref{sec:Markov}. A summary of our results and an outlook to future work are provided in Section \ref{sec:SumCon}. Details on the stochastic convection parameterisations can be found in Appendices~\ref{app.RV} and \ref{app.MC}.


\section{Data}
\label{sec:Data}
\subsection{Description of the datasets of tropical convection in Kwajalein and Darwin}
\label{sec:Data1}
We utilise two datasets of observations of the large-scale vertical velocity at 500~hPa $\omega_{500}$ and of the concurring ${\rm CAFs}$ and rain rates over tropical locations, averaged to yield 6-hourly time resolution. The datasets each cover a $190\times190$\,km$^2$ pentagon-shaped area centered over Darwin (Australia) and Kwajalein (Marshall Islands), respectively. The area is chosen as to represent the size of a typical climate model grid-box. The Kwajalein site is located in the tropical western Pacific and is typical for a purely tropical oceanic climate. The Darwin site on the other hand is typical for the monsoon climate of northern Australia and features the complex topography characteristic of a coastal site. 

The area-mean values of atmospheric variables are derived using the method of \cite{Xie04}, who employ the variational analysis approach of \cite{Zhang97}, but use profiles of atmospheric variables from numerical weather prediction models instead of atmospheric soundings. Here, the variational analysis employs analyses from ECMWF and is constrained by observations of surface precipitation obtained from C-band polarimetric (CPOL) research radars \citep{Keenan98} and top-of-the-atmosphere radiation at both locations to reliably balance the column budgets of mass, heat, moisture and momentum. \cite{Davies13} show that constraining the variational analysis by observed rainfall substantially improves the derived large-scale vertical velocities over the Darwin domain compared to using just the ECMWF analysis alone.

Over Darwin, the analysis is applied to observational data obtained during three consecutive wet seasons (2004/2005, 2005/2006, 2006/2007), yielding a total of 1890 6-hour means. Over Kwajalein, the analysis is applied to the time period of May 2008 -- Jan 2009, produced to fit into the framework of the Year Of Tropical Convection virtual field campaign \citep{Waliser08,Waliser12}. For Kwajalein, 1095 6-hour means are available. At both locations, the large-scale atmospheric data are complemented by data of the concurrent small-scale convective state derived from CPOL radar observations. The radar observations were used to derive rain area fractions attributable to either stratiform or convective precipitation after \cite{Steiner95}. The convective area fraction is then determined as the ratio of the number of radar pixels classified as ``deep convective'' with respect to the total number of pixels. More information regarding the derivation of the datasets can be found in \cite{Davies13}. 

By relying on available 6-hourly averaged data, some characteristics of tropical convection, e.g.~the diurnal cycle, are ill-resolved. The advantage of the 6-hourly averaged data used in this study is that they are self-consistent in the sense that the large-scale state is determined via the variational analysis and constrained by the radar observations to satisfy budgets of mass, heat, moisture, and momentum using the variational analysis \citep{Davies13}. We are not aware of observational data with higher temporal resolution with the same properties covering a comparable time period.

The data have already provided important new insights into the behaviour of tropical convection \citep{Davies13,Peters13,Kumar13}. In particular, \cite{Peters13} showed that the relationship between convection and a range of large-scale atmospheric forcing conditions is very similar for both regions despite their distinctly different atmospheric and oceanic regimes.


\subsection{Analysis of the datasets over Kwajalein and Darwin}
\label{sec:Data2}
To support our premise that the underlying stochastic process relating the small-scale convective activity to the large-scale variables is sufficiently independent of the geographical location, we contrast here the observed convection at Darwin and Kwajalein.

%
\begin{figure*}[htbp]
\centering
\includegraphics[width = \columnwidth]{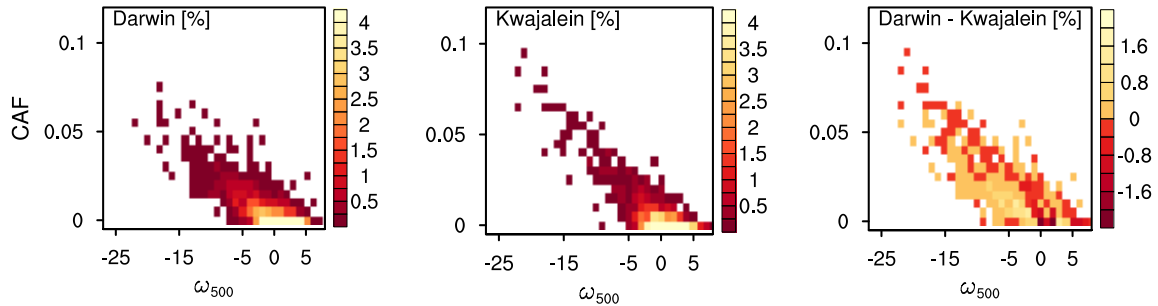}
\caption{Normalised $2d$ histograms of ${\rm CAF}$ and  $\omega_{500}$ [hPa/hour] obtained from observations over Darwin (left) and Kwajalein (middle). The difference of the histograms is depicted in the right most plot.}
\label{fig:DataAll2DHistos}
\end{figure*}

Figure~\ref{fig:DataAll2DHistos} shows the $2d$ histograms of ${\rm CAF}$ and $\omega_{\rm 500}$ of the observations in Darwin and Kwajalein as well as the difference between the two distributions. Throughout the paper $\omega_{500}$ is given in units of [hPa hour$^{-1}$]. The plots show strong qualitative similarities between the two locations which are suggestive of the existence of a universal relationship which can be utilised to construct stochastic subgrid-scale parameterisations of ${\rm CAF}$ conditioned on the large-scale variable $\omega_{500}$. Let us briefly discuss some of the particularities of the relationships between ${\rm CAF}$ and $\omega_{\rm 500}$ in Kwajalein and Darwin, as seen in Figure~\ref{fig:DataAll2DHistos}. The difference between the distributions (right panel in Figure~\ref{fig:DataAll2DHistos}) shows that Darwin features more convective activity in the range of $-5< \omega_{500} < 0$ than Kwajalein. The converse is true for $0< \omega_{500} < 5$. We attribute this difference in convective behaviour to the different prevailing meteorological conditions  in Kwajalein and Darwin and the respective different convection initiating mechanisms. In particular, land-sea breeze induced convective organisation at Darwin (diurnal cycle), and the generally more inhomogeneous surface characteristics of the Darwin domain may contribute to different convective responses given a particular large-scale forcing. For relatively weak large-scale dynamical forcing,  i.e.~for $-5<\omega_{500}<5$ in our case, land-surface heterogeneities in the Darwin region, such as coastlines or spatial differences in land cover, can induce subgrid-scale mesoscale circulations leading to organised convection \cite[e.g.][]{Pielke01,Rieck14} which then results in increased mean large-scale ascent. The increased convective activity in Darwin for negative values of $\omega_{500}$ implies a concurrent decrease for positive values in the histograms as seen in Fig.~\ref{fig:DataAll2DHistos} due to the normalisation. 
It is worth mentioning that the observations also include several instances of zero precipitation; $236$ events from $1890$ observations in Darwin and $28$ from $1095$ observations in Kwajalein, and several instances of zero CAF; $194$ such events in Darwin and $82$ in Kwajalein. Note that a zero CAF does not imply that there is no precipitation and vice versa. Significant deep convection is possible for neutral or even mean subsiding conditions as in, for example, land-sea breezes in the tropics during mean suppressed conditions.

Further, Figure~\ref{fig:DataAll2DHistos} shows that the variance of ${\rm CAF}$ is dependent on the state $\omega_{500}$ and increases with decreasing values of $\omega_{500}$ (not shown). This is consistent with the result of \cite{CraigCohen06} and \cite{CohenCraig06} that the variance of convective activity increases with the forcing. Therein the forcing considered was a range of radiative cooling rates. However, we remark that, increased radiative cooling is typically compensated by increased domain mean mass flux, and therefore the vertical velocity $\omega_{500}$ is an effective proxy for forcing. \cite{Peters13} show that the ratio of the standard deviation and the mean of ${\rm CAF}$ decreases for sufficiently negative values of $\omega_{500}$. This suggests that heavy rain events may be viewed as being deterministic (relative to weaker rain events) with an approximate linear dependency on $\omega_{500}$. This is particularly evident in the Kwajalein data (Figure~\ref{fig:DataAll2DHistos}, middle panel). An analysis of coarse-grained outputs from the ECMWF IFS shows similar results for the Darwin region \citep{Watson15}.

\begin{figure}[t!]
\centering
\includegraphics[width = 0.5\columnwidth]{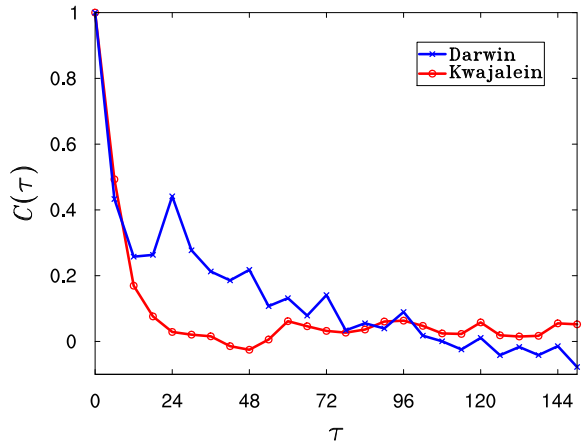}
\caption{Temporal autocorrelation $C(\tau)$, with $\tau$ in hours, of the ${\rm CAF}$ time series for Darwin (blue crosses) and Kwajalein (red circles).}
\label{fig:ACFs}
\end{figure}

Figure~\ref{fig:ACFs} shows that ${\rm CAF}$ observed at Kwajalein and Darwin has similar autocorrelation up to lags of 12 hours. For lags longer than 12 hours, convection over Kwajalein looses memory, whereas convection over Darwin exhibits significant autocorrelation up to lags of 72 hours and features peaks corresponding to the diurnal cycle (every 24 hours).

%


\subsection{Statistical similarity of convective activity}
\label{sec:universal}

The comparison of convective behaviour in Darwin and Kwajalein above suggests that both locations feature notably different convective behaviour. In this Section we will nevertheless establish crucial similarities in the relationship between convective activity and large-scale vertical motion which constitute the working hypothesis for our stochastic parameterisation schemes. A reader who is just interested in the actual stochastic parametrisation may skip this section upon first reading.\\

The stochastic subgrid-scale parameterisations proposed in the next Section utilise conditional probabilities such as $p({\rm CAF}(t)|\omega_{\rm 500}(t))$ describing the probability of convective activity ${\rm CAF}$ occurring at time $t$ for given vertical velocity $\omega_{\rm 500}$ at that time. We therefore now compare empirical conditional probabilities for the two locations, Darwin and Kwajalein, which we denote by $p_{\rm{Darwin}}$ and $p_{\rm{Kwajalein}}$, respectively. To construct the conditional probabilities we bin the $(\omega_{\rm 500},{\rm CAF})$-domain into bins of size $(0.1,0.01)$.

Assuming that the different prevailing atmospheric and oceanic regimes impact directly on the large-scale variables, we consider as a first approximation a uniform translation of the large-scale vertical velocities. In particular, we show that the conditional probability functions $p_{\rm{Darwin}}$ and $p_{\rm{Kwajalein}}$ are close when the vertical velocities of Darwin are shifted as in
\begin{eqnarray}
&p^{\rm Kwajalein}({\rm CAF}(t)|\omega_{\rm 500}(t)) \approx \qquad \qquad \qquad \qquad\nonumber \\
& \qquad \qquad \qquad p^{\rm Darwin}({\rm CAF}(t)|\omega_{\rm 500}(t) - \Delta_\omega)
\end{eqnarray}
 or analogously 
 \begin{eqnarray}
&p^{\rm Darwin}({\rm CAF}(t)|\omega_{\rm 500}(t))  \approx \qquad \qquad \qquad \qquad\nonumber \\
& \qquad \qquad \qquad  p^{\rm Kwajalein}({\rm CAF}(t)|\omega_{\rm 500}(t) + \Delta_\omega)
\; .
\end{eqnarray}

\begin{figure}[htbp]
\centering
\includegraphics[width = 0.5\columnwidth]{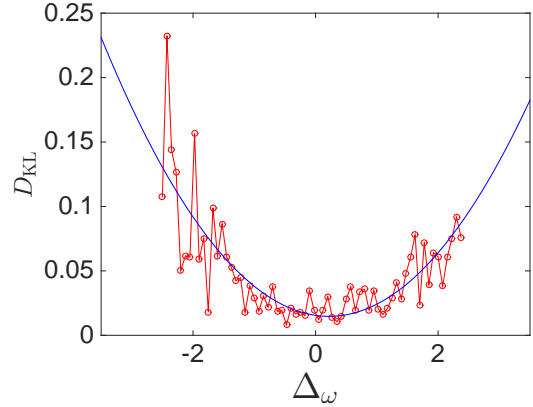}
\caption{Kullback-Leibler distance between the conditional probability functions $p_{\rm{Darwin}}$ and $p_{\rm{Kwajalein}}$ as a function the shift $\Delta_\omega$ (circles). The minimum of the quadratic least square approximation (solid curve) is at $\Delta_\omega=0.21$.}
\label{fig:DKL}	
\end{figure}

A standard tool to compare probability density functions $P$ and $Q$ is their Kullback-Leibler distance
\begin{eqnarray}\label{e.KL}
&D_{\rm KL}(P||Q) =\int \log\left(\frac{P(x)}{Q(x)} \right) \, P(x) \,dx\; .
\end{eqnarray}
The Kullback-Leibler distance $D_{\rm KL}(P||Q)$ is defined provided that the support of the probability function $P$ is contained in the support of $Q$; otherwise it is infinite. The Kullback-Leibler distance is a non-negative quantity and it is zero if and only if $P=Q$ (see for example \citet{Kantz}).\\
We will estimate the Kullback-Leibler distance between the conditional probabilities $p_{\rm Kwajalein}$ and $p_{\rm Darwin}$ for each of the $\omega_{\rm 500}$-bins. In Figure~\ref{fig:DKL} we show the median of these Kullback-Leibler distances $D_{\rm KL}$ as a function of the global shift $\Delta_\omega$. We have discarded those $\omega_{\rm 500}$-bins for which the support of the conditional probability for Darwin is not contained in the support of that for Kwajalein to allow for finite values of $D_{\rm KL}$.

%

A quadratic regression yields an optimal shift of $\Delta_\omega = 0.21$ where the minimum of the Kullback-Leibler distance is attained. The shift $\Delta_\omega$ is given in units of [hPa hour$^{-1}$]. In general, the Kullback-Leibler distance is asymmetric with $D_{\rm KL}(P||Q) \neq D_{\rm KL}(P||Q)$. We find, however, that $D_{\rm KL}(p_{\rm{Kwajalein}} ||  p_{\rm{Darwin}})$ has a minimum very close to same value of $\Delta_\omega$ supporting our approximation that the two conditional probability functions are related by a simple translation of the vertical velocities. We note, that due to the larger amount of available observations for Darwin ($N=1890$) when compared to Kwajalein ($N=1095$) and due to the larger support of $p_{\rm{Kwajalein}}$ the formulation (\ref{e.KL}) is preferred.\\

\begin{figure}[htbp]
\centering
\includegraphics[width = 0.5\columnwidth]{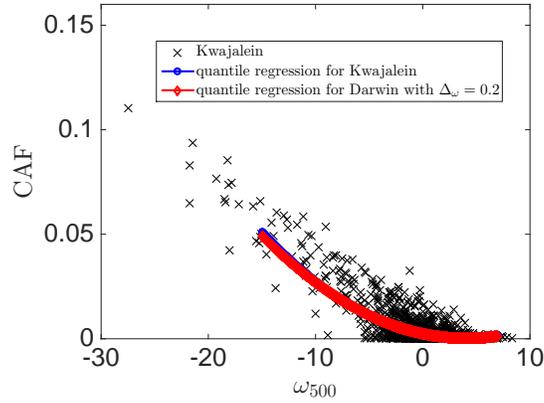}
\caption{${\rm CAF}$ as a function of the vertical velocities $\omega_{500}$ [hPa/hour] obtained from observations over Kwajalein (black crosses). The continuous line connecting the circles (online blue) shows the results of a $2^{\rm{nd}}$order median regression. The continuous line connecting the diamonds (online red) shows the result of a $2^{\rm{nd}}$order median regression for the Darwin data plotted against $\omega_{500}-0.2$.}
\label{fig:QR}
\end{figure}

The similaritiy of the convective behaviour at both locations can be further examined by performing a median (or $50^{\rm{th}}$-quantile) regression for ${\rm CAF}$ (see for example \cite{KoenekerBassett78,QR} and \citet{Bremnes04,FriederichsHense08,Mudelsee} for applications of quantile regression methods in the atmospheric sciences). We determine the conditional median for the observations of Kwajalein and Darwin using a second-order regression. Using conditional medians rather than conditional means (as done in normal least square regression) produces more robust estimates by eliminating the impact of the few very large rain events and other statistical outliers. The median regressions for Kwajalein and for Darwin approximately coincide if one translates the $\omega_{500}$ values of Kwajalein by $\Delta_\omega = 0.2$ (or those of Darwin by $-\Delta_\omega=-0.2$, respectively), as seen in Figure~\ref{fig:QR}, corroborating the finding of the Kullback-Leibler analysis.\\

We remark that the shift $\Delta_\omega$ depends on the height at which $\omega$ is evaluated. We also analysed observations of the vertical velocity taken at 715~hPa; there the optimal shift for which the respective quantile regressions were closest and for which the Kullback-Leibler distance was minimal is found to be $\Delta_\omega \approx 1.67$. We attribute this uniform shift of the large-scale vertical velocity to the different prevailing atmospheric-oceanic regimes at the two respective locations as discussed in Section~\ref{sec:Data2}. Specifically, land surface effects are expected to exert a stronger influence on atmospheric variables in the lower (715~hPa) than in the middle (500~hPa) troposphere.\\

It is by no means clear that the same (possibly non-zero) shift can be applied to all locations in the tropics. The particular value of $\Delta \omega$ found from the data in Darwin and Kwajalein might be different when considering other geographical locations. Furthermore, it is also not clear that a similarity of the conditional probability functions exists at all when shifting the vertical velocity for other geographical locations. This would have to be checked when more data from other locations become available. If true, such a universality would mean that no costly geographically dependent fine-tuning would be required in estimating the shift $\Delta_\omega$ for different geographical locations.\\

The estimated shift $\Delta_\omega=0.2$~hPa hour$^{-1}$, we found here for $\omega_{\rm 500}$, is small compared to the range of $\omega_{\rm 500}$ and we therefore ignore the shift when comparing data from Kwajalein with Darwin (and vice versa), unless stated otherwise (note that non-trivial shifts have to be applied in constructing models conditioned on, let's say, $\omega_{\rm 715}$). To ensure sufficient generality, however, we will present in the following Section the method for possible non-trivial shifts $\Delta_\omega \neq 0$.



%


\section{Stochastic subgridscale parameterisation}
\label{sec:Markov}

We will develop two stochastic subgrid-scale parameterisation schemes for ${\rm CAF}$ conditioned on $\omega_{500}$; one in which subgrid-scale convection variables such as ${\rm CAF}$ are viewed as instantaneous random variables conditioned on the current value of the large-scale vertical velocity $\omega_{500}$, and a second approach in which the subgrid-scale variables are viewed as a conditional Markov chain taking into account non-vanishing temporal correlations of the subgrid-scale variables (cf.~Figure~\ref{fig:ACFs}). The para\-metrisation schemes we propose model tropical convection at any location given only the information of the large-scale values of $\omega_{500}$ at a given time without any usage of the small-scale convection variables such as ${\rm CAF}$ at that time.\\ 

We are given time series consisting of 6-hourly averaged observations of $\omega_{500}$ and of ${\rm CAF}$ obtained at Kwajalein and Darwin, which we denote by $\{\omega_{{500}_k}\}_{k=1,\cdots,N}$ and $\{y_k\}_{k=1,\cdots,N}$ with $N=1890$ for Darwin and $N=1095$ for Kwajalein, respectively (cf. Section \ref{sec:Data}). The statistical similarity of convective activity established in Section~\ref{sec:universal} suggests that we can generate the stochastic model from observations of either location and apply it to the other location, respectively, without applying a linear shift $\Delta_\omega$ to the vertical velocities $\omega_{500}$. We describe the methods for the situation when observations obtained in Darwin are used to train the model which is then subsequently applied to observations of $\omega_{500}$ in Kwajalein, but we will present results as well for the reversed case. 


\subsection{Instantaneous conditional random variables}
\label{sec:RV}
In our first stochastic model convective activity is treated as sequence of independent random variables conditioned on the current value of the vertical velocity $\omega_{\rm 500}$. The parameterisation has two components: a training component and an application component. The training component is performed as follows. Given pairs of observations for the vertical velocity $\omega_{500}$ and ${\rm CAF}$ (or the rain rate), we want to associate with each value of $\omega_{500}$ a range of possible convective events and determine their respective probabilities of occurrence. We do so by partitioning the $(\omega_{500},{\rm CAF})$-plane into bins. This will define coarse-grained values $\hat \omega$ for $\omega$. For each of the coarse-grained values $\hat \omega$ we can now associated coarse-grained values $\widehat{\rm CAF}$ by averaging ${\rm CAF}$ over each bin associated with the coarse-grained value $\hat \omega$ and estimate their respective conditional probabilities $P(\widehat{\rm CAF}|\hat \omega)$ empirically recording the frequencies of $\widehat{\rm CAF}$ in their respective bins. The interested reader is referred to the Appendix~\ref{app.RV} for more details on the model.

We construct the stochastic model with observations from Darwin. We partition the $(\omega_{\rm 500},y)$-plane into bins of size $(0.8,0.005)$. 
Choosing the bin size is a balancing act between requiring sufficiently small bin sizes to assure accuracy and needing sufficiently large bin sizes to allow for meaningful statistical averages within a bin. Choosing the bins requires tuning and is dependent on the number of observations available. We have tested that doubling the bin sizes still produces good results.\\

Since we do not have sufficient data to construct the stochastic model for large negative values of $\omega_{500}$, we use a deterministic relationship between ${\rm CAF}$ and $\omega_{\rm{500}}$ for observations with $\omega_{\rm 500}<-18$ (cf. \citet{Peters13}). The deterministic relationship is found by linear regression of the observations to be ${\rm CAF}=-0.0044\; \omega_{\rm 500}-0.011$.\\

\begin{figure}[t]
\centering
\includegraphics[width = 0.5\columnwidth]{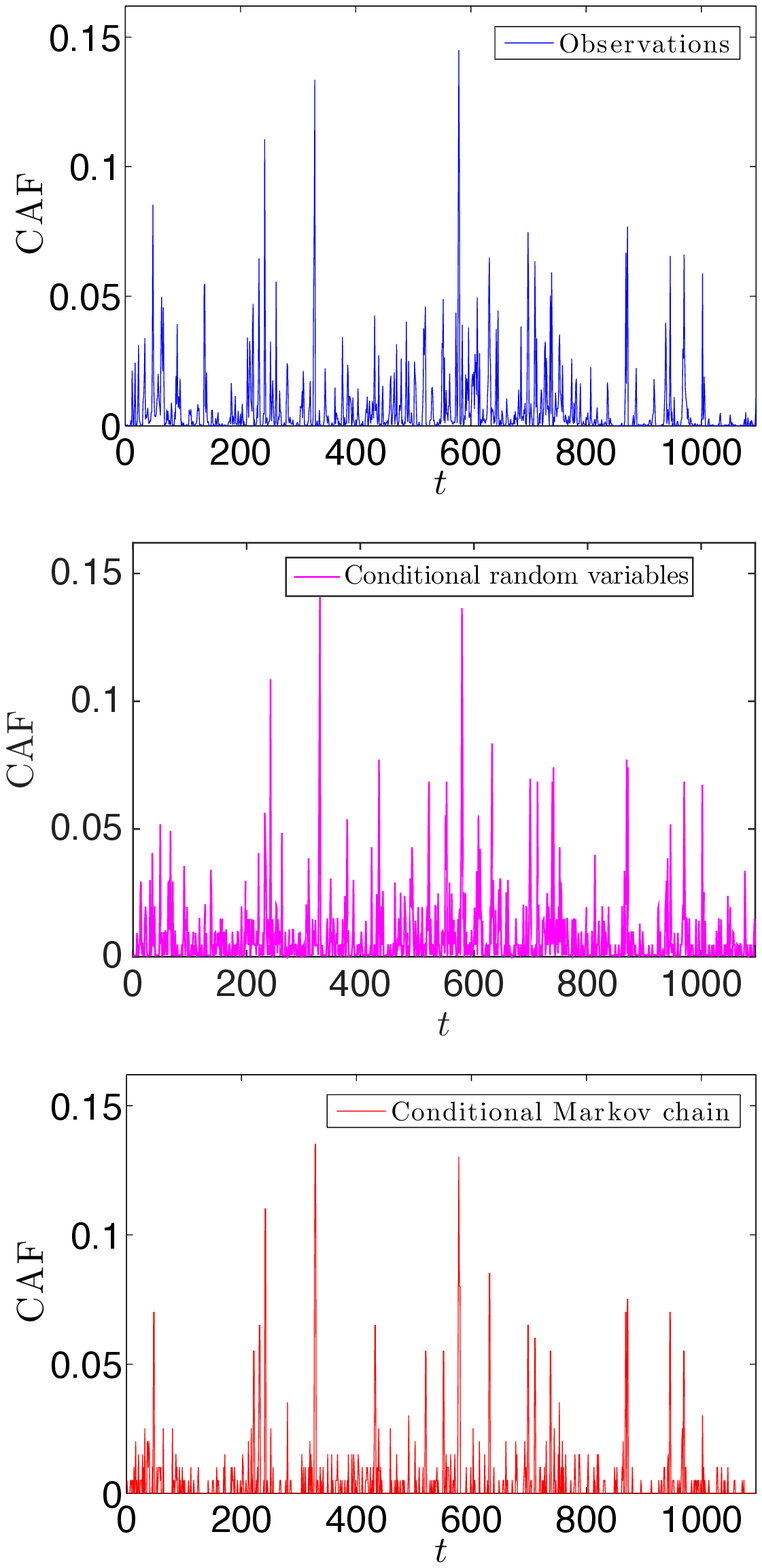} 
\caption{Time series of ${\rm CAF}$ of the observations over Kwajalein (top), of the synthetic process conditioned on the vertical velocities $\omega_{500}$ described in Section~\ref{sec:RV} (middle) and of the conditional Markov process process described in Section~\ref{sec:CMC} (bottom). The time series generated via the conditional Markov chain has missing data points in the depicted time interval (see text for details). The plots have a time resolution of 6 hours. Here $t=0$ corresponds to 1 May 2008 00 UTC.}
\label{fig:timeseries}
\end{figure}

To test the effectiveness of our model we now apply the Darwin-trained model to observations in Kwajalein and generate synthetic time series of ${\rm CAF}$ conditioned on the large-scale $\omega_{\rm 500}$ observed over Kwajalein. In Figure~\ref{fig:timeseries} we show the time series of the observations of ${\rm CAF}$ in Kwajalein (top panel) and the corresponding synthetic time series of the stochastic model using conditional random variables (middle panel). The model reproduces observed intermittent features of tropical convection. However, it fails to reproduce periods of sustained non-convection near, for example, $t\approx 200$ and $t\approx 900$. This failure is due to our approach not incorporating any memory or trends, despite non vanishing autocorrelations as seen in Figure~\ref{fig:ACFs}. 

\begin{figure}[htbp]
\centering
\includegraphics[width = 0.5\columnwidth]{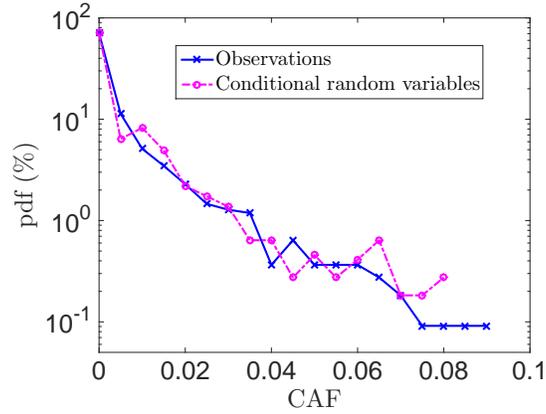}
\caption{Empirical histogram of ${\rm CAF}$ for the observations over Kwajalein (crosses, online blue) and for the synthetic process conditioned on the vertical velocities $\omega_{500}$ described in Section~\ref{sec:RV} (circles, online magenta).}
\label{fig:pdf}
\end{figure}

To establish a more quantitative comparison, we compare in Figure~\ref{fig:pdf} the empirically determined probability density functions of ${\rm CAF}$ for the synthetic time series and the actual observations. By performing averages over $1,000$ realisations of the stochastic model we have established that the first three moments of ${\rm CAF}$ in Kwajalein, the mean $\mu$, the variance $\sigma^2$ and the skewness $\xi$, are well captured by our synthetic time series. This is illustrated in Table~\ref{table:ConvAreaFrac}. \\

\begin{table}[htbp]
\caption{First three moments mean $\mu$, variance $\sigma^2$  and skewness $\xi$ of observed ${\rm CAF}$ for Kwajalein and of the synthetic data obtained by the subgrid-scale parameterisations conditioned on $\omega_{500}$ for the two models trained with observations from Darwin.}
\label{table:ConvAreaFrac}
\begin{center}
\begin{tabular}{cccc}
  &  $\mu$ & $\sigma^2$ & $\xi$\\
\hline
observations & $0.0066$ & $1.89\, 10^{-4}$ & $4.27$\\
random variable &  $0.0073$ & $1.80\, 10^{-4}$ & $4.29$\\
Markov chain &   $0.0066$ & $2.75\, 10^{-4}$ & $4.25$\\
\hline
\end{tabular}
\end{center}
\end{table}
\begin{table}[htbp]
\caption{First three moments mean $\mu$, variance $\sigma^2$  and skewness $\xi$ of observed ${\rm CAF}$ for Darwin and of the synthetic data obtained by the subgrid-scale parameterisation conditioned on $\omega_{500}$ for the two models trained with observations from Kwajalein.}
\label{table:ConvAreaFrac_K_to_D}
\begin{center}
\begin{tabular}{cccc}
  &  $\mu$ & $\sigma^2$ & $\xi$\\
\hline
observations & $0.0080$ & $1.29\, 10^{-4}$ & $2.38$\\
random variable &   $0.0075$ & $1.45\, 10^{-4}$ & $2.46$\\
Markov chain &   $0.0083$ & $2.38\, 10^{-4}$ & $2.46$\\
\hline
\end{tabular}
\end{center}
\end{table}

The numerical results presented above used a stochastic model which was generated using the observations at Darwin and then subsequently applied to observations of large-scale vertical velocities observed at Kwajalein to produce the associated convective activity at Kwajalein. In accordance with the statistical similarity of convective activity established in Section~\ref{sec:universal} we have also trained the stochastic model on the data observed at Kwajalein and applied them to observations of large-scale vertical velocities observed at Darwin with equal success. The results for the first three moments are shown in Table~\ref{table:ConvAreaFrac_K_to_D} for completeness.

We remark that conditioning the observations on the large-scale variables produces better estimates of the moments than simply taking the observations. For example, the actually observed mean of convective activity in Kwajalein $\mu=0.0066$ is estimated as $0.0073$ using instantaneous random variables conditioned on $\omega_{500}$ (cf. Table~\ref{table:ConvAreaFrac}) whereas if just estimated by the mean of the training set (i.e. the observations of CAF in Darwin) the estimate of the mean of convective activity would be $0.008$ (cf. Table~\ref{table:ConvAreaFrac_K_to_D}). 

We obtained similarly good results when parameterising ${\rm CAF}$ conditioned on observations of the vertical velocity at 715~hPa (not shown); in this case the vertical velocities were shifted by $\Delta_\omega=1.67$ (cf. Section~\ref{sec:universal}). 

Further, we have constructed synthetic time series of rain rate data consisting of random variables conditioned on the vertical velocity and found similarly good results (not shown).


\subsection{Conditional Markov chain}
\label{sec:CMC}

The observational data obtained in Kwajalein and Darwin exhibit non-vanishing temporal autocorrelations as illustrated in Figure~\ref{fig:ACFs}. This suggests that a more appropriate parameterisation of ${\rm CAF}$ should incorporate dependencies on previous observations rather than simply conditioning on the present values of the large-scale variables. The autocorrelation of ${\rm CAF}$ and of $\omega_{500}$ as well as of the crosscorrelation function for a lag of one time step (6 hours) exhibit similar values in Kwajalein and in Darwin (with the autocorrelation function for $\omega_{500}$ exhibiting a much stronger diurnal cycle), but differ substantially for lags greater than 12 hours. This suggests a Markov model trained at one location should adequately capture the convective behaviour at the other location if conditioned on only the observations of the previous time step 6 hours ago. As a first step towards incorporating memory one may construct a Markov chain conditioned on the previous state of the system (see, for example, \citet{CrommelinVandenEijnden08}) or by fitting an AR(1) process about an $\omega_{500}$-dependent mean as in \citet{Wilks05}. We follow here the approach proposed by \citet{CrommelinVandenEijnden08} for a conditional Markov chain. The conditional Markov chain estimates the conditional transition probability $P(\widehat{\rm CAF}_k|\hat \omega_k,\hat \omega_{k-1},\widehat{\rm CAF}_{k-1})$, where $k$ denotes the present time and $k-1$ the time of the previous observation. The conditioning on the previous time step takes into account trends in the dynamics of the vertical velocity and accounts for non-trivial temporal correlations. We first estimate the (unconditional) transition probability from observations at Darwin. This is achieved again by partitioning the $(\omega_{500},{\rm CAF})$-plane into bins and counting frequencies of transitions between bins within one sampling time. The aim is now to use this transition probability to draw random realisations $\widehat{\rm CAF}_k$ from this Markov chain for observations $(\hat \omega_{k-1},\widehat{\rm CAF}_{k-1})$ in Kwajalein conditioned on the current observation of the large-scale velocity $\hat\omega_k$. We refer the interested reader to the Appendix~\ref{app.MC} for more details on practical aspects.

The data sparse region of large convective activity for $\omega_{\rm 500}<-18$ is again treated with a deterministic relationship as in the instantaneous random variable model described in Section~\ref{sec:RV}. 
We subdivide the $(\omega_{\rm 500},{\rm CAF})$-plane again into bins of size $(0.8,0.005)$.

In Figure~\ref{fig:timeseries} (bottom panel) we show a time series of the observations of ${\rm CAF}$ in Kwajalein and the corresponding data obtained from the conditional Markov chains which was trained with observations obtained in Darwin. Due to insufficient amount of data not all transitions could be captured leading to a shorter synthetic time series. Only approximately $3/4$ of the data points in Kwajalein can be reached by the Markov chain and only approximately $60\%$ of those form a time-continuous set of at least $12$ hours. Hence the plot of the time series in Figure~\ref{fig:timeseries} suffers from missing data points along the given time interval. We mention that \cite{Dorrestijn15} employed a Markov chain model for the data obtained in Darwin mitigating the problem of data sparseness by i), coarse-graining the convective state into different cloud types at the scale of individual radar pixels, rather than using ${\rm CAF}$ directly, and ii) using precipitation area fraction data at very high temporal resolution (10 minutes) in combination with a linearly interpolated version of the 6-hourly large scale atmospheric state. We chose not to employ such a linearly interpolated version of the large-scale data as this eliminates the self-consistency of the dataset.

\begin{figure}[htbp]
\centering
\includegraphics[width = 0.5\columnwidth]{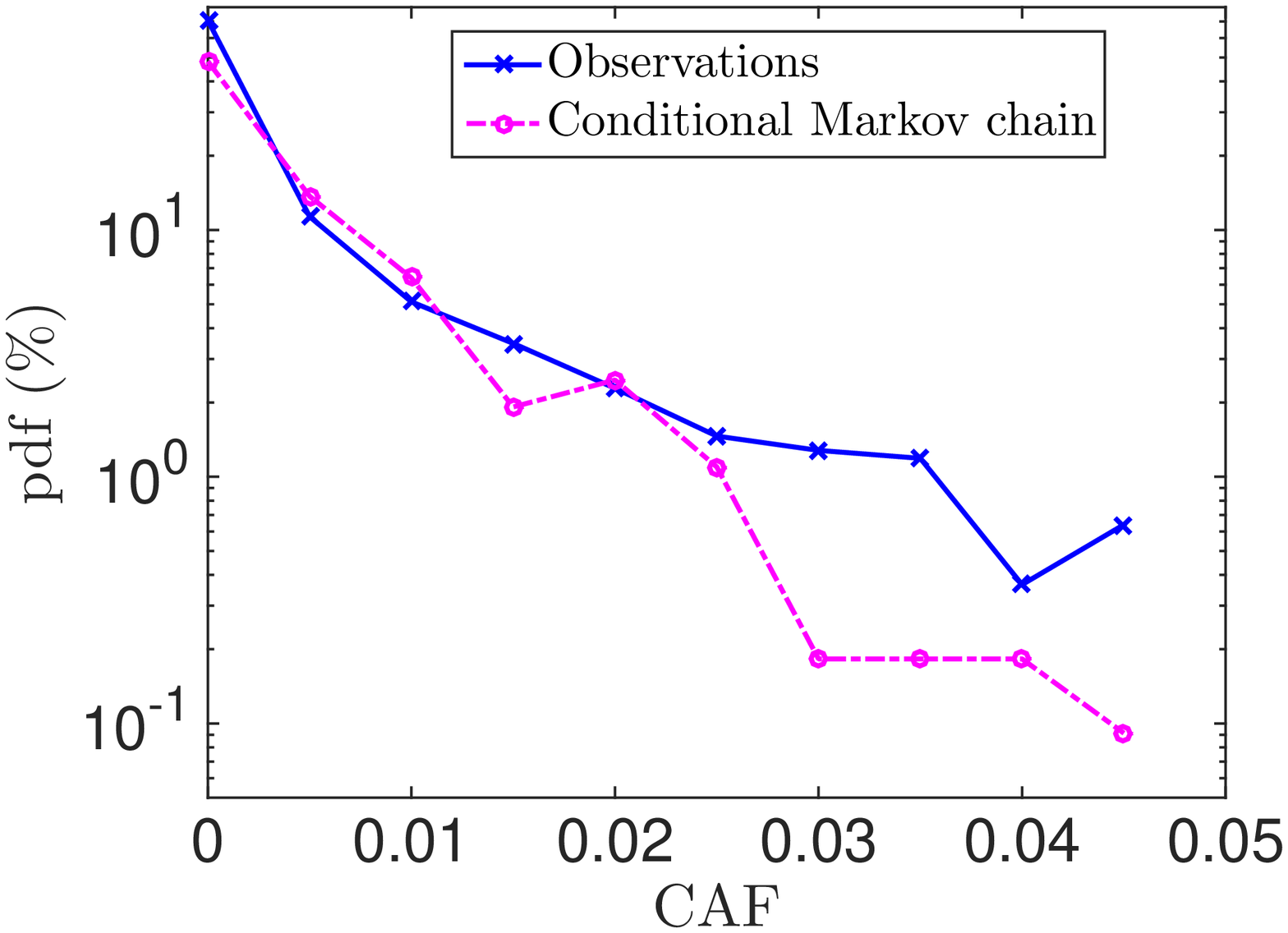}
\caption{Empirical probability density function of ${\rm CAF}$ for the observations over Kwajalein (crosses, online blue) and for the conditional Markov chain model (circles, online magenta).}
\label{fig:pdfCMC}
\end{figure}

The empirical probability density functions of ${\rm CAF}$ are shown in Figure~\ref{fig:pdfCMC} with reasonable correspondence. Results 
for the first three moments of ${\rm CAF}$ are listed in Tables~\ref{table:ConvAreaFrac} and \ref{table:ConvAreaFrac_K_to_D}. Again, the statistics of the actual observations is reasonably well reproduced. The variance is overestimated by the Markov chain. This may be due to the averaging of ${\rm CAF}$ within the relatively coarse bins (cf. the definition of the coarse-grained ${\rm CAF}$ values (\ref{ybar}) which is also used in the Markov chain). 

%


\section{Summary and Conclusions}
\label{sec:SumCon}
In this study, we used observations of tropical deep convection and the concurring large-scale atmospheric states at two tropical locations, Darwin and Kwajalein, to design a data-driven stochastic subgrid-scale parameterisation for tropical deep convection. The parameterisation we propose can be built off-line and then subsequently implemented at low computational cost. The schemes we proposed assume that convective activity has been triggered.\\


Given large-scale variables such as vertical velocity, as provided by the dynamical core of the host model, our stochastic models can be coupled to an already existing convection scheme, which is part of the model physics. The important and hard problem of triggering convection is performed by the host models' convection scheme. Once convection is triggered, we see the contribution of our stochastic models as providing the host models' convection scheme with statistically consistent estimates for the cloud-base mass flux. Properly estimating the cloud base mass flux is paramount to determine the overall strength of convection. This can be done in a scaleable way using CAF to determine the convective cloud base mass flux. The convective cloud base mass flux can be estimated as the product of CAF, the air density and the upward velocity at cloud base which may be either assumed constant, e.g.~1 ms$^{-1}$, or may be estimated from boundary layer characteristics. The upward velocity at cloud base would be assigned at the beginning of the updraught calculation in the convection scheme, with CAF providing the link to the large-scale environment.\\ 

We presented two diagnostic approaches to stochastically parameterise convective activity conditioned on large-scale vertical velocity. The first method treated ${\rm CAF}$ as an instantaneous random variable conditioned on the current value of $\omega_{\rm 500}$. This method suffers from neglecting non-vanishing autocorrelations present in the observations and is not able to reproduce periods of sustained convection and non-convection, for example. The second approach was built around a conditional Markov chain and incorporates autocorrelations to some degree; this method, however, requires substantially more data to train the Markov chain as it involves conditioning on the past observations as well as on the current value of $\omega_{\rm 500}$. Given these limitations, our results are promising. The marginal probability functions of ${\rm CAF}$ as well as its first three moments were reasonably well reproduced by both approaches, except for the variance which was overestimated by the Markov chain. This is particularly remarkable as the stochastic models were trained with data from one geographical location and then applied to another geographical location with different atmospheric and oceanic conditions. 
In general, we would expect the conditional Markov chain to provide better diagnostics than the parameterisation consisting of instantaneous random variables as it accounts for memory effects. In particular we expect the conditional Markov chain to reproduce the autocorrelations of ${\rm CAF}$ for time lags less than 6 hours. The Markov chain generated by our observational data sets, however, did not produce long enough artificial time series of ${\rm CAF}$ which would allow for a reliable estimation of the autocorrelation function. 
To further test the proposed parameterisation schemes for ${\rm CAF}$ we will in future work i) use numerical data from high-resolution cloud resolving models (or larger observational data sets if they become available) and ii) implement the proposed stochastic models as part of operational convection parameterisations in comprehensive GCMs. 

We have used quantile regression and the Kullback-Leibler test to probe for universality of the relationship between convective activity and large-scale vertical motion at 500~hPa, $\omega_{500}$ [hPa/hour], allowing for a simple global shift of the vertical velocities. Despite markedly different prevalent atmospheric and oceanic regimes at Darwin and Kwajalein the joint probability density functions were close and did not require a shift. This implied that the stochastic models can be trained at one geographical location and then be subsequently applied to the respective other location. For vertical velocities evaluated at 715~hPa the joint probability density functions were closest, however, when a simple shift in the vertical velocity was performed. To more accurately calibrate the required shifts in the vertical velocities and to take into account the respective atmospheric environments of different geographical locations, numerical data from high-resolution cloud resolving models could be used as a surrogate for missing observational data in future research.\\

We chose to parameterise mainly subgrid-scale ${\rm CAF}$ because i) it is directly related to domain mean rainfall and thus total latent heating and ii) assigning a non-zero area fraction to convective updrafts in a convection scheme relieves the problems associated with the assumption of ``scale-separation'' as employed in current convection schemes \cite[e.g.~][]{Arakawa11}. 
As described above, our stochastic models could be efficiently applied to estimate statistically consistent estimates of cloud base mass flux, essentially providing the closure for mass-flux convection schemes. 
Such a convective scheme would be fully scalable with convective updrafts eventually covering large portions of or even entire grid-boxes. In fact, ongoing work by one of the authors (KP) shows that such an implementation yields plausible results in a full GCM. Although ${\rm CAF}$ is suited for a resolution independent comprehensive parameterisation of deep convection, the way the observational data have been obtained involves a particular spatial scale (i.e. the $190\times190$\,km$^2$ pentagon-shaped area considered here). The observations would have to be adapted for the particular resolution of the GCM. In that context, using the same data as in the present study, \cite{Tikotin12} sub-divided the Kwajalein domain into sub-domains of different size and analysed the relationship between convective activity and $\omega_{500}$ as a function of the domain size. While the overall statistical relationships remain identical with decreasing domain size, the variability of convection given a particular large-scale state increased with decreasing domain size.\\

We have developed here stochastic parameterisation schemes for convective activity which are data-driven. Their attractiveness lies in their simplicity and their ease of implementation. They can be a useful tool in times when the physics is not sufficiently well understood and/or resolved by physics-based parameterisation schemes. However, we would like to end with a word of caution for data-driven parameterisation schemes in climate models. The models are trained under the assumption of statistical equilibrium. It is not clear whether the change of global climatic conditions will leave the statistical relationships between ${\rm CAF}$ and $\omega_{500}$ constant. These issues would not apply to parametrisation in numerical weather prediction models.\\


In climate or numerical weather prediction models the resolved variables including the vertical velocities are updated in time using the convective state, e.g. vertically resolved heating rates. To be able to test whether our data-based stochastic parametrisation for the convective state can be successfully used requires several tests planned for further research. Our premise is that, given a judicious choice of large-scale variable, convective activity can be parametrised in terms of just these variables. In this work we chose the vertical velocity at $500$ hPa as our large-scale variable. The stochastic models we proposed here are only practically viable if the number of those judicious variables is sufficiently small. 
This is similar to the approach taken by \cite{Dorrestijn15} who conditioned a data-driven stochastic multi-cloud model on large-scale vertical velocity only and were able to adequately simulate observed convective area fractions.
Of course, the strength of atmospheric moist convection also depends on numerous other variables such as the buoyancy of surface air parcels and humidity of the mid-troposphere. It is {\em{a priori}} not clear whether conditioning on just one variable is sufficient. Indeed, one could imagine that by neglecting the conditioning of the convective state on more variables than just the large-scale vertical velocity, the error in the stochastic parametrisation for the convection will eventually be accrued in all large-scale variables during the numerical integration. This may lead to a detrimental accumulation of errors in a positive feedback loop. 
It is planned to test in high-resolution cloud resolving models whether introducing more than on large-scale variable for the conditioning will be beneficial. In particular, low-to mid level moisture might be important as it is known to play a major role in, for example, in the initiation of the Madden-Julian Oscillation; see for example,  \cite{KhouiderEtAl13,AjayamohanEtAl13} and references therein. 
In case more resolved large-scale variables are needed to condition the stochastic parametrisations of convective activity, one could use a linear combination of these variables to allow for a computationally feasible parametrisation.


\paragraph{Acknowledgements}
GAG and KP acknowledge support from the Australian Research Council. We thank Garth Tarr and Neville Weber for discussions on quantile regression. We thank Steven Sherwood, Bob Plant, Chris Holloway, Glenn Shutts and two anonymous referees for constructive comments and suggestions on an earlier version of the paper.


\appendix

\section{Description of the stochastic model using instantaneous random variables}
\label{app.RV}

Let us denote by $y$ the subgrid-scale variable, for example ${\rm CAF}$ or the rain rate. We partition the range of $\omega_{500}$ into $N_{\omega}$ intervals $I^i_\omega$ with $i=1,\cdots,N_\omega$ and the range of the subgrid-scale variables into $N_y$ intervals $I_y^n$ with $n=1,\cdots,N_y$. This partitions the $(\omega_{500},y)$-plane into $N_\omega N_y$ bins. We assume that the time series $\{\omega_{{500}_k}\}_{k=1,\cdots,N}$ and $\{y_k\}_{k=1,\cdots,N}$ stem from a stationary process. Coarse-grained ${\rm CAF}$ values (denoted by $\widehat{\rm CAF}$ in Section~\ref{sec:RV}), conditioned on the large-scale variables $\omega_{500}\in I_\omega^i$ (denoted by $\hat \omega$ in Section~\ref{sec:RV}), are determined as averages over bins with
\begin{equation}
\label{ybar}
\bar y^{(n,i)} = \frac{\sum_k y_k {{\bf 1}[y_k\in I_y^n]\, {\bf 1}[\omega_{{500}_k}}\in I_\omega^i]}{N_y^{(n,i)}}\, ,
\end{equation}
where $N_y^{(n,i)}=\textstyle{\sum_k {{\bf 1}[y_k\in I_y^n] \, {\bf 1}[\omega_{{500}_k}}\in I_\omega^i]}$ is the number of $y_k$-values belonging to the bin defined as the intersection of the intervals $I_\omega^i$ and $I_y^n$. Here ${{\bf 1}}[\cdot]$ denotes the indicator function with ${{\bf 1}}[y_k\in I_y^n]=1$ if $y_k\in I_y^n$ and ${{\bf 1}}[y_k\in I_y^n]=0$ otherwise. The conditional probability $P(n|i)$ of ${\rm CAF}$ $y_k$ being in the interval $I_y^n$ conditioned on $\omega_{{500}_k}$ being in the interval $I_\omega^i$ (denoted by $P(\widehat{\rm CAF}|\hat \omega)$ in Section~\ref{sec:RV}) is calculated as
\begin{equation}
P(n|i) = \frac{\sum_k {\bf 1}[y_k\in I_y^n]\, {\bf 1}[\omega_{{500}_k}\in I_\omega^i]}{N_y^i}\; ,
\label{Pni}
\end{equation}
where $N_y^i=\textstyle{\sum_k {\bf 1}[\omega_{{500}_k}\in I_\omega^i]}$ is the number of realisations of $y_k$ for a given value of the large-scale $\omega_{{500}_k}\in I_\omega^i$. Note that $\sum_n P(n|i)=1$. Estimating $\bar y^{(n,i)}$ and $P(n|i)$ concludes the training period. 

To generate artificial time series of the subgrid-scale variable $y$ conditioned on $\omega_{500}$ observed at a different geographical location, one simply assigns with probability $P(n|i)$ the coarse grained value ${\bar y}^{(n,i)}$.


\section{Description of the stochastic model using a conditional Markov chain}
\label{app.MC}
To construct the Markov chain we determine a transition probability $P_{i,n}^{j,m}$ which denotes the probability for the variables $(\omega_{{500}_k},y_k)$ to take values in the bin defined as the intersection of the intervals $I_\omega^j$ and $I_y^m$ at time step $k$ when they were in the bin defined as the intersection of the intervals $I_\omega^i$ and $I_y^n$ at the previous time step $k-1$. To construct $P_{i,n}^{j,m}$ as a matrix we arrange the bins into one long array. The associated $N_\omega N_y\times N_\omega N_y$ transition matrix $P_\alpha^\beta$ describing transitions from bin $\alpha=i+(n-1)N_\omega$ to bin $\beta=j+(m-1)N_\omega$ is then estimated from the observational data as
\begin{equation}
P_{\alpha}^{\beta} = \frac{T_{\alpha}^{\beta}}{\sum_{\beta=1}^{N_\omega N_y} T_{\alpha}^{\beta}} \, ,
\label{Pij}
\end{equation}
where $T_{\alpha}^{\beta}$ counts the number of transitions from the bin labelled with $\alpha$ to the bin labelled with $\beta$ and is given by
\begin{equation}
  \begin{split}
    \nonumber
    T_{\alpha}^{\beta} &= \sum_k {\bf 1}[\omega_{{500}_{k-1}}\in I_\omega^i]\, {\bf 1}[y_{k-1}\in I_y^n]\\
    & \quad\;\;\; \times 
    {\bf 1}[\omega_{{500}_{k}}\in I_\omega^j]\, {\bf 1}[y_{k}\in I_y^m]\; .
    \label{Tij}
  \end{split}
\end{equation}

Estimating the transition matrix $P_{\alpha}^{\beta}$ concludes the training phase. To construct a Markov chain conditioned on $\omega_{500}$ taking a particular value at present time step $k$, we apply the transition matrix to the given past state $\alpha^\star$ at time $k-1$ to calculate $\pi_{\alpha^\star}^{\beta} = (0,\cdots,1,\cdots,0)P_{\alpha}^{\beta}$ where the $1$ is in the $\alpha^\star$-th entry. Then we select those $L \leq N_y$ bins, i.e. the non-zero coordinates of $\pi_{\alpha^\star}^{\beta}$, which are consistent with the current value $\omega_{{500}_k}$. These $L$ entries of $\pi_{\alpha^\star}^{\beta_l}$ with $l=1,\cdots,L$, associated with the current value of $\omega_{500}$, (if they exist!), do not necessarily sum up to $1$ as required for a probability. Hence we renormalise as follows 
\begin{equation}
\tilde \pi_{\alpha^\star}^{\beta_l} = \frac{\pi_{\alpha^\star}^{\beta_l}}{\sum_{l=1}^L \pi_{\alpha^\star}^{\beta_l}}\, .
\label{pi}
\end{equation}
The subgrid-scale variable $y_k$ is then randomly chosen from $L$ possible states with probability $\tilde \pi_{\alpha^\star}^{\beta_l}$. The assigned values corresponding to the bin labelled with $\beta_l$ are coarse-grained values obtained by averaging over the bins analogously to (\ref{ybar}). 



\end{document}